**Title**
The paradoxical relationship of difficulty and lateral frontal cortex activity.

**Authors**
Christopher H. Chatham[1], Nicole M. Long[2], and David Badre[3,4]

**Affiliations**

[1]Clinical & Translational Imaging, Pfizer Worldwide Research & Development
[2]Department of Psychology, University of Pennsylvania
[3]Cognitive, Linguistic, and Psychological Sciences, Brown University
[4]Brown Institute for Brain Science, Brown University

**Competing financial interests**
The authors declare no competing financial interests.

**Corresponding authors**
Christopher Chatham (chathach@gmail.com)
David Badre (david_badre@brown.edu)

**Running Title**
Rostrocaudal reaction time effects

**Abstract**

Task difficulty is widely cited in current theory regarding cognitive control and fronto-parietal function. Ongoing debate surrounds the extent to which global difficulty across multiple cognitive demands is the main driver of lateral frontal activity. Here, we examine a commonly cited behavioral marker of difficulty in these accounts: time-on-task (ToT), as assessed by response time. Specifically, we investigate the task-dependent scaling of frontal BOLD responses with ToT during hierarchical cognitive control. We observe a paradoxical relationship, whereby rostral regions show greater scaling with ToT on a first-order task, despite showing greater recruitment on a second-order task; caudal regions show the converse relationships. Together, these results demonstrate that ToT does not reflect a single dimension of difficulty that uniformly drives lateral frontal activity. Rather, this discrepancy in the mean and scaling of BOLD requires that multiple, distinct processes are instantiated across these fronto-parietal regions in the service of cognitive control function.

**Main Text**

Cognitive control refers to the array of mechanisms enabling goal-directed behavior when habits prove insufficient. Difficult tasks often require deliberative, strategic, and goal-directed processing. Therefore, cognitive control is generally more necessary as tasks become more difficult. Most prior research in cognitive control has sought to fractionate difficulty into subtypes relying on dissociable neurocognitive mechanisms. Some of the posited subtypes pertain to the multifaceted origins of task difficulty, such as stimulus- vs. response-related conflict[1] or temporal or rule abstraction[2-5]. Orthogonal characterizations involve the temporal dynamics by which such difficulties are addressed, e.g., via proactive vs. reactive[6-7] or serial vs. parallel decision making[8-10]. Yet another approach involves dissociating closely intertwined sources of individual differences in batteries of difficult tasks, such as between shifting vs. updating abilities[11-13].

However, recent evidence suggests that a unitary construct of task difficulty may also play an important role in the mechanisms of cognitive control. This proposal is motivated by three diverse sources of evidence. First, a large network of frontoparietal regions – sometimes referred to as the "multiple demand" or "frontoparietal control" network – responds with surprising uniformity during difficult tasks, despite the heterogenous subcomponents of task difficulty enumerated above and the anatomical heterogeneity in the frontoparietal network itself[14-15]. Second, task difficulty itself might require a unified cognitive representation, to permit the evaluation of whether a task is too demanding to be worth performing[16-17]. Third, there are quite general demands imposed by all cognitive control tasks, to which individuals respond in highly characteristic and even heritable ways[12]. This global ability to perform well on difficult tasks may be at the basis of fluid intelligence[15]. Thus, there are a variety of empirical and normative reasons to expect that task difficulty may not be entirely reducible to separable neurocognitive subcomponents.

To test a unitary construct of difficulty, difficulty must be measured in multiple ways. This requirement stems from the fact that even perfect double-dissociations in a single dependent variable can be explained by single-process models[18-21]. By contrast, the use of multiple dependent variables enables a more constrained test: any mapping from a single latent process to two (or more) dependent variables also implicitly determines a mapping between the dependent variables themselves. For example, if task difficulty is assessed in more ways than mere "time-on-task" (i.e., reaction time; RT) – e.g., if it is also assessed in terms of the frontoparietal BOLD response – then a mapping from BOLD to RT is also thereby posited (Figure 1A). This mapping can then be used to test, and potentially reject, a large class of single-process models that relate difficulty to both time-on-task and frontoparietal recruitment (Figure 1B-C). Thus, the delineation of relationships among dependent variables, and in particular that between RT and BOLD, is crucial for testing what a unitary difficulty construct alone can explain in the domain of cognitive control.

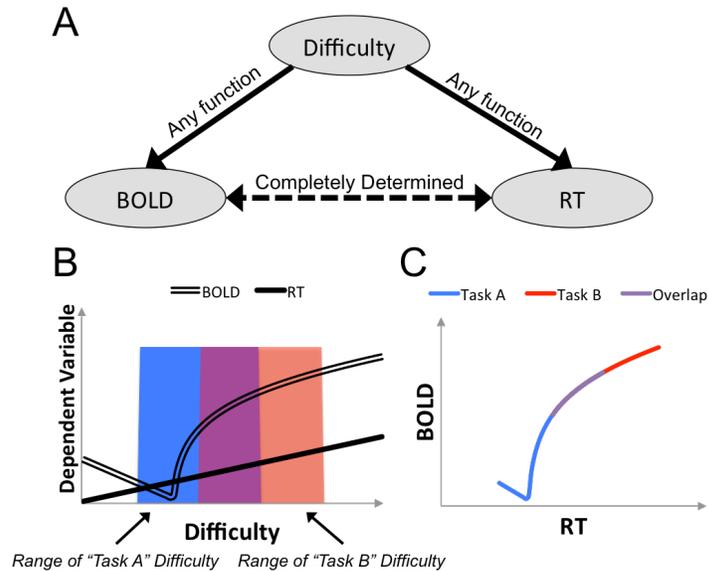

**Figure 1 – Multiple dependent variables render single-process models falsifiable**. Classical dissociation logic and more recent state-trace methods cannot distinguish single from multiple process models on the basis of a single dependent measure[18-21] However, the use of multiple dependent measures can allow for the rejection of certain single-process models. **A.** If the difference between two tasks in two dependent variables (here, BOLD and RT) is to be explained on the basis of a latent single process (here, "difficulty"), then one must specify two mappings: the mapping between difficulty and BOLD, and between difficulty and RT. However, these mappings by necessity also make a prediction for the relationship between the dependent measures themselves. **B.** To illustrate, consider tasks with partially-overlapping ranges of difficulty: "Task A", which is generally somewhat less difficult (blue range), and "Task B", which is generally somewhat more difficult (red range). Any differences in BOLD and RT between two tasks can always be explained by drawing arbitrarily complex curves relating difficulty to them (black lines). Implicitly, however, this also specifies a predicted relationship of BOLD to RT that can be tested empirically (**C**). For example, the RT/BOLD relationship should be identical across tasks that are matched for RT (purple region), if RT is a monotonic function of difficulty, regardless of monotonicity in the mapping from difficulty to BOLD.

This agenda is complicated by a host of rarely-mentioned methodological issues. These issues generally derive from the fact that RT is itself a dependent measure, and hence cannot be treated the same as other explanatory variables in a general linear model (GLM). First, while the GLM returns unbiased parameter estimates despite errors in dependent variables, it is biased by errors in the independent variables (via regression dilution[22]). While experimental factors can be imposed without error by an experimenter, RT does contain measurement error. As a result, RT effects are uniquely subject to regression dilution when used in a traditional GLM for fMRI. Second, whereas experimental effects are typically optimized by design to yield frequency spectra where the BOLD signal is most sensitive, RTs intrinsically vary with the 1/f spectrum dominated by MR noise[23]. As a result, RT/BOLD mappings may vary between tasks due to differences in the frequency spectra of the RTs they elicit, rather than true differences in the underlying mapping from RT to neural activity. Finally, differences in the observed range of RTs in different conditions may produce artifactual differences in RT/BOLD mappings across those conditions, if the actual RT/BOLD mapping is nonlinear (e.g., Fig 1C). To our knowledge, none of these issues has been addressed in prior investigations of RT/BOLD relationships, despite the widespread use of "brain-behavior" correlations in cognitive neuroscience, and the ongoing debate regarding the kind and number of latent constructs necessary to explain cognitive control-related effects in RT and BOLD[24-29].

Here, we address these challenges to examine whether a global "difficulty" construct, as reflected in RT, can explain frontal BOLD responses typically associated with cognitive control. Specifically, we administered to subjects a concrete, first-order task (a so-called "1D" task involving only one conditionality, of the form "if shape A, then button X; if shape B, then button Y"; see Figure 2A) as well as a more abstract, second-order task (a "2D" task involving a second level of conditionality, of the form: "if color 1, then respond according to the shape rules; if color 2, then respond according to the texture rules"; see Figure 2B). Multiple prior studies show that more abstract, higher order tasks, like the 2D task, drive the recruitment of more rostral regions in prefrontal cortex[2-5], but task difficulty has also been offered as an alternative explanation for these results[14; c.f. 30]. Our results will show a paradoxical relationship between RT and BOLD in fronto-parietal cortex suggesting that task difficulty alone cannot be considered a unitary explanation of heterogeneity in the frontal lobe BOLD response to demands on cognitive control.

**Figure 2. Task Design.** Following an instructed training phase in which subjects learned to respond with one of four fingers to each of four shapes and textures (not illustrated; see Methods), subjects completed interleaved blocks of two tasks while undergoing fMRI. (**A**) In blocks of the 1D task, subjects were first cued as to which of the dimension rules they had practiced (shape vs. texture) would be relevant for that block. Color was thus entirely irrelevant. (**B**) In blocks of the 2D task, subjects were cued as to which colors mapped to which dimension. Subjects therefore had to use color as a context for adjudicating between two policies – responding on the basis of shape, vs. responding on the basis of color.

## Results

*Behavioral Results – Error Rates.* By the final block of training, subjects performed with a mean error rate (ER) of 4.1% (range 0-11%), indicating they understood the rules and had memorized the feature-to-response mappings. Overall accuracy during the experimental phase was likewise excellent (mean ER 7%; range: 0-13%).

A 2 (task: 1D vs. 2D) x 2 (frequency: high vs. low) repeated measures analysis of variance (RM-ANOVA) on ER revealed increased errors in the 2D task (F(1,21)=100.87, p<.001; Figure 3A).

*Behavioral Results – RT*. The D1 and D2 tasks elicited several changes in the distribution of reaction time. Correct trial RT was tested using a 2 (task: 1D vs. 2D) x 2 (frequency: high vs. low) RM-ANOVA. RT was increased on the 2D task (F(1,21)=1033.34, p<.001; Figure 3B). Further effects of task on the RT distribution were revealed by plotting the mean RT within each decile of the condition-specific RT distribution (Figure 3C), and subjected to statistical analysis using within-subject measures of RT standard deviation, skew, and kurtosis. In particular, RT's standard deviation (Figure 3D) was increased in the 2D condition (F(1,21)=53.70, p<.001). The 2D task was also less positively skewed than the 1D task (F(1,21)=182.83, p<.001; Figure 3E) and showed reduced kurtosis (F(1,21)=48.67, p<.001; Figure 3F). In summary, the 1D and 2D tasks are characterized by an array of differences in the RT distribution, with significant task effects present in each of its first four moments.

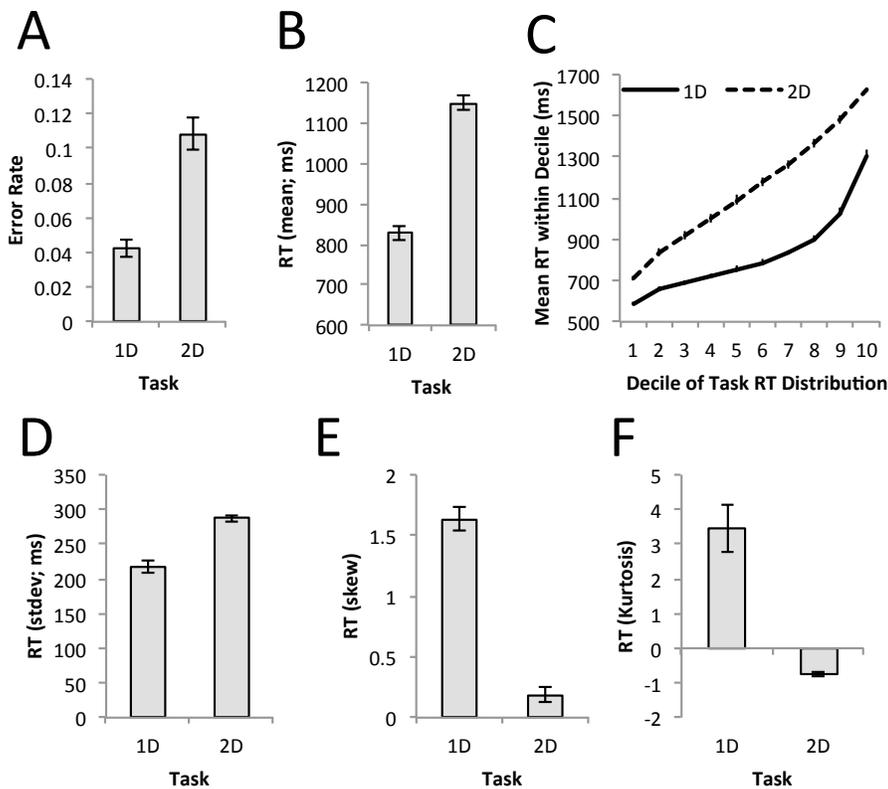

**Figure 3. Task-related differences in behavior.** Performance of the 2D task differed from the 1D task in several ways, including elevated error rates (**A**), increases in mean RT (**B**), and several further alterations to the RT distribution, as displayed in the decile plot of (**C**). In particular, the 2D task was associated with increased RT variability (as assessed by the within-subject standard deviation of RT; **D**), decreased skew (**E**), and reduced kurtosis (**F**).

*Imaging Results – Main Effects of Task*. The 1D and 2D tasks were both associated with significant BOLD change, relative to fixation, in a wide set of frontoparietal regions (Figure 4A; black outlined regions). Further, an overlapping set of frontal, parietal, and occipital sites showed a significantly greater BOLD response to 2D task events than 1D task events (Figure 4A; orange regions). Conversely, regions typically associated with the "default mode" network were less deactivated by 1D task events, and hence showed greater BOLD during 1D than 2D task events, including the medial prefrontal cortex, middle temporal lobe, and the temporoparietal junction (Figure 4; blue regions).

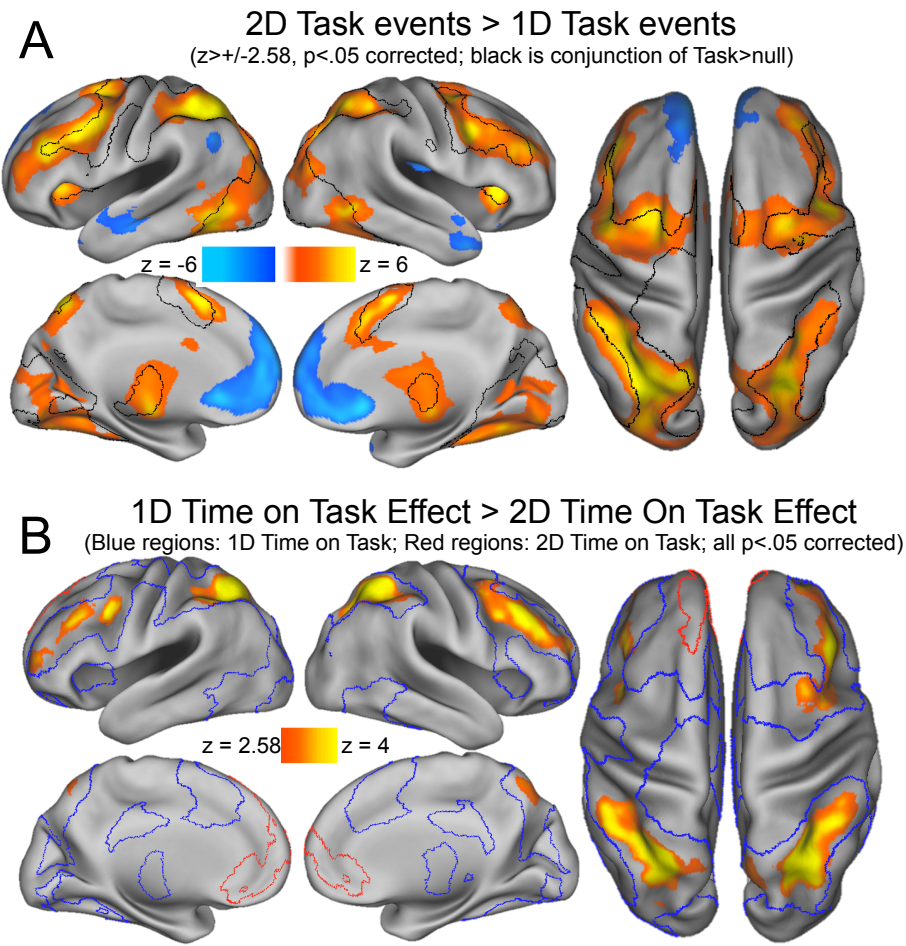

**Figure 4. Mean frontoparietal BOLD to 1D and 2D tasks (A) and its scaling with RT (B). A.** Though the 1D and 2D task alike yielded reliable activation in the variety of frontal, parietal, occipital and subcortical sites typical of demands on cognitive control (black outlined regions represent the conjunction across tasks), the 2D task was nonetheless associated with a significantly greater BOLD response in most of these areas (orange regions). The 1D task was associated with a significantly greater BOLD response in a more restricted set of areas, including temporo-parietal, posterior insula, middle temporal, and both anterior medial and midline prefrontal regions. **B.** Blue outlines show regions exhibiting a significant parametric response as a function of time spent on the 1D task; red outlines show regions with a significant parametric response to time spent on the 2D task. These time-on-task effects were significantly stronger in the 1D task in specific foci within rostrolateral prefrontal cortex, inferior frontal sulcus, and the dorsal pre-premotor cortex, along with a broad swath of the intraparietal sulcus, its dorsal bank, and the superior parietal lobule. No regions showed a significantly greater scaling with time on the 2D task after correction for multiple comparisons.

*Imaging Results – Effects of Time-on-task*. To assess the hemodynamic consequences of the different reaction times observed in the 1D and 2D tasks (see Figure 3), we added two variable-epoch regressors to the GLM, corresponding to correct RTs in the 1D and 2D tasks respectively. As can be seen in Figure 5, time-on-task effects in a variety of frontoparietal regions were reliably stronger within the 1D task, relative to the 2D task (orange regions of Figure 4B). Indeed, time on the 1D task related to activity in a large set of frontal, parietal and occipital regions (blue outlines of Figure 4B) whereas BOLD scaled with time on the 2D task solely in midline prefrontal regions generally associated with the default mode.

As introduced above, differences in RT/BOLD scaling between conditions can arise from a number of sources. For example, differences in parameter estimates can arise from the use of independent variables with variable error (via regression dilution[22]). However, the split-half reliability of RT was excellent on both tasks (Cronbach's Alpha of .946 and .935 for odd vs. even RTs in the 1D and 2D tasks, respectively; Fig 5A), indicating that regression dilution could not explain the present results. Alternatively, differences in our parameter estimates might have arisen due to differences in the frequency spectrum of our regressors. For example, any regressor whose spectral power lies within the frequency bands typically removed from the BOLD timeseries (such as through high-pass filtering) would appear to be associated with less BOLD signal change. However, the convolved RT regressor of the 2D task actually had greater power across all frequency bands relevant to the BOLD response than the 1D task. In particular, the power of the 2D task's RT regressor peaked at roughly the same frequency as the BOLD response, implying that if differences in spectral profile were the only factor driving the relation to RT, then the 2D task should actually have been associated with greater BOLD/RT scaling than the 1D task, not less (Figure 5B).

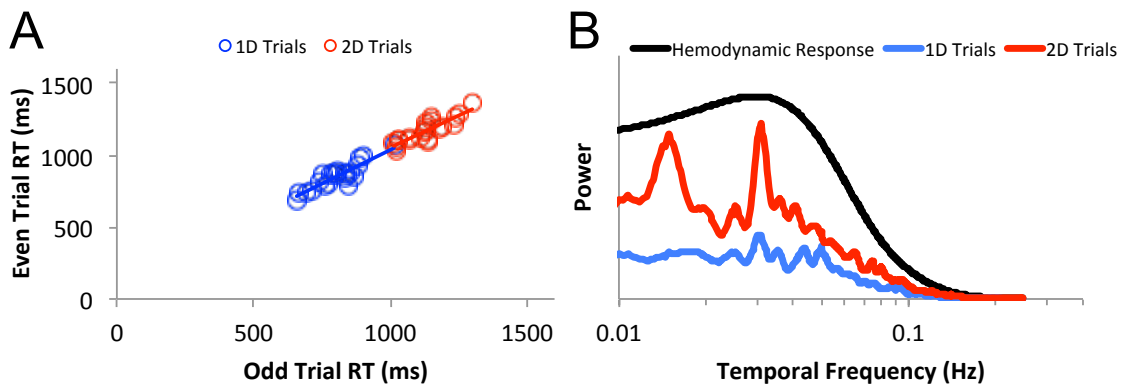

Figure 5. **Task differences in the RT/BOLD mapping cannot be explained by regression dilution (A) or differences in filter-matching with the BOLD response (B)**. **A.** In the GLM, parameter estimates are biased in proportion to error in explanatory variables. However, the split half reliability of the tasks was excellent, suggesting that differences in measurement error could not explain the differential effects of time-on-task between 1D and 2D tasks. **B.** Though experimental factors are typically optimized to match the spectrum of the BOLD response (black line), RT is not under direct experimental control. As a result, artifactual differences in the BOLD correlates of RT could reflect differences in the spectral power. However, we find that the 2D task is associated with greater power across the spectrum relevant to the BOLD response, with a prominent peak in the vicinity of the peak power of the canonical hemodynamic response itself. Thus, all else being equal, the 2D task should have been associated with greater measured BOLD change – contrary to observation.

We next used a multivariate matching algorithm to subsample the 1D and 2D task RT distributions so as to construct two sets of task events that were matched not only in terms of RT, but also all experimental variables other than task itself (see Methods). Despite the multifaceted differences in the original RT distribution across the two tasks (left subpanel of Figure 6A), our procedure generated a set of trials matched in the first four moments of the distribution both at the aggregate (right subpanel of Figure 7A) and individual subject levels (Figure 6B). Variable-epoch regressors were then constructed on the basis of these matched RTs. Despite this close matching of RT, frontoparietal regions continued to show a stronger scaling with RT in the 1D, as compared to 2D task (Figure 7C, orange regions). A greater scaling of BOLD with RT in the 1D, as compared to 2D task was limited to regions that showed a significant scaling with matched RTs in the 1D task (blue outlines in Figure 6C), and not the 2D task (red outlines in Figure 6C).

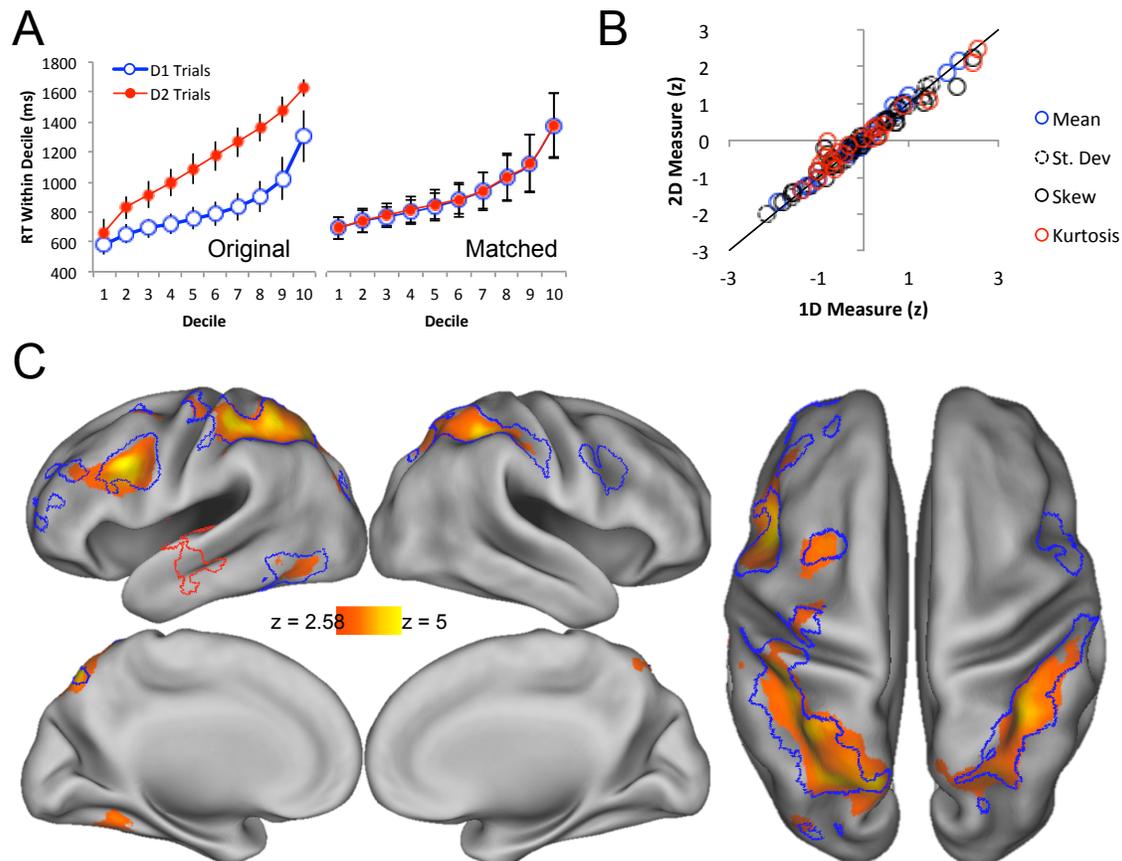

**Figure 6. The temporal scaling of the frontoparietal BOLD response was significantly greater in the 1D task even when RT and other trial characteristics were precisely matched across tasks. A.** A genetic multivariate matching algorithm was used to subsample the task-specific RT distributions and to construct a subset of trials that were precisely balanced across tasks (see Methods). **B.** This matching algorithm yielded RT distributions matched across tasks for each individual separately, in terms of the first four moments of the RT distribution. To illustrate the success of this approach, we standardized the mean, standard deviation, skew and

kurtosis of each individual's 1D and 2D task RT distributions with respect to the other instances of that same measure across all other individuals and tasks. Thus, not only was the RT distribution matched in aggregate across tasks, but individual differences in each aspect of the distributions were also matched across tasks. **C.** Outlined blue regions show bilateral prePMd and parietal cortex as well as left RLPFC and IFS regions scale with 1D task RTs, whereas 2D task RTs yield a significant BOLD response in middle temporal cortex. Left prePMd/IFS and bilateral parietal regions showed a prominently increased response to time spent on the 1D task, relative to matched periods of time spent on the 2D task.

Next, we constructed a "maximal model" controlling for other experimental factors that might differ by task. This model controls for the effects of stimulus congruency, of cue and response switching, and of sustained block-related effects. The differential scaling of BOLD with RT during the 1D task remained significant in select bilateral frontoparietal regions across all trials, including those previously associated with hierarchical control[3] (RLPFC, white region, MNI: +/-36, 50, 8; IFS, blue region, MNI: +/- 52, 24, 24; and PrePMd, green region; PMd is also shown, in red, MNI: +/-32, -12, 66; Figure 7A). At the whole-brain level the left Pre-PMd (and adjacent IFS) continued to show a significantly greater scaling of BOLD with matched trials in the 1D, relative to 2D task in this "maximal" model (Figure 7B; see also Table S1).

A region-of-interest analysis of the *a priori* Pre-PMd confirmed that the scaling of bilateral Pre-PMd BOLD with RT was larger during matched 1D trials (as compared to 2D; $F(1,21)=4.57$, $p=.04$). Conversely, the mean BOLD response in this area was nevertheless larger on average during matched 2D trials (as compared to 1D; $F(1,21)=4.83$, $p<.04$). These two patterns were significantly different from one another ($F(1,21)=9.78$, $p=.005$), and were unlike those observed in the more caudal PMd (task [1D vs. 2D] x measure [height of constant duration vs. height of variable-epoch regressor] x area [Pre-PMd vs. PMd] interaction: $F(1,21)=5.17$, $p=.03$).

We next sought to characterize task differences in how RT/BOLD scaling influences the shape of the hemodynamic response using a finite impulse response (FIR) model in each of the frontal ROIs previously associated with hierarchical control (colored regions in Figure7A&B). These analyses revealed a rostrocaudal dissociation in which more rostral areas were more associated with time spent on the more concrete 1D task, rather than the more abstract 2D task. In this analysis, the most caudal region, PMd, showed a larger scaling with time-on-task in the 2D task bilaterally; by contrast, more rostral regions showed a larger scaling with time-on-task in the 1D task bilaterally (Figure 7C). An RM-ANOVA of average BOLD PSC associated with matched RTs in the 4-8 seconds following stimulus onset yielded a reliable interaction in the focused contrast of area [PMd vs. more rostral areas] and task [1D vs. 2D] ($F(1,21)=6.51$, $p=.019$).

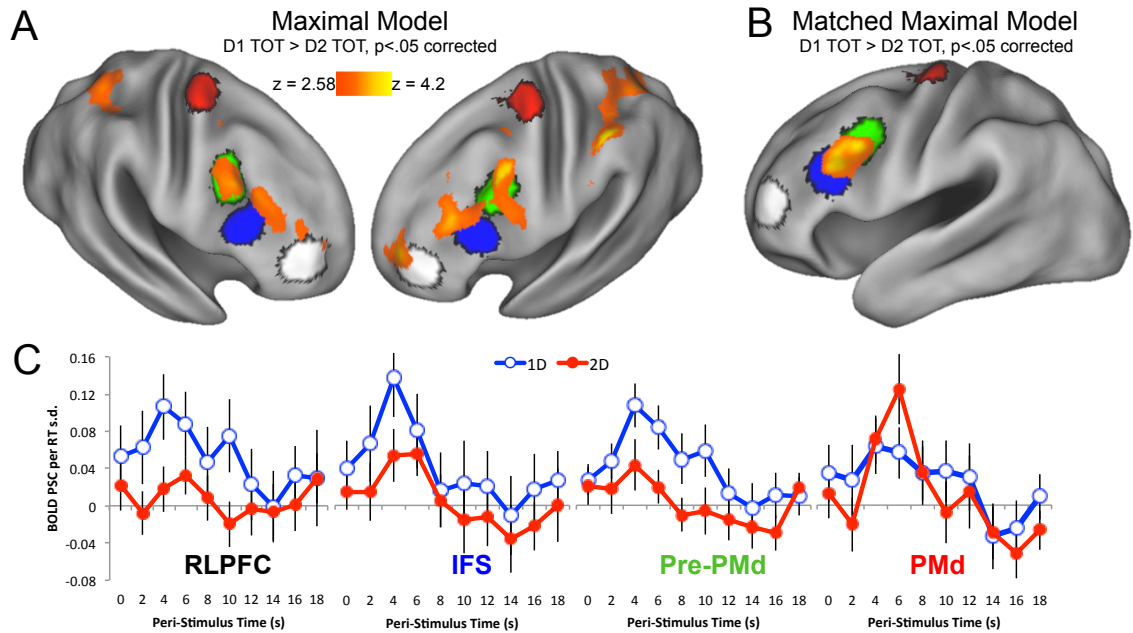

**Figure 7. Rostrocaudal effects reverse with time-on-task. A**. Greater scaling of BOLD with RT was observed in the 1D, as compared to 2D task even when controlling for a variety of experimental factors. The resulting foci bear a striking similarity to the foci previously implicated in more abstract forms of cognitive control, including the RLPFC (white region), IFS (blue regions), Pre-PMd (green regions). No results survived whole brain correction in PMd (red) region most strongly associated with first-order tasks . **B.** Greater scaling of BOLD with RT in the 1D task remained significant in the IFS and PrePMd when considering matched trials only, indicating the effect in these rostral regions cannot be attributed to the different ranges of RT observed across those tasks. **C**. A volume-by-volume analyses of RT/BOLD scaling (via an FIR model; see Methods) yielded a three-way interaction between rostral extent, task, and time-on-task, such that RT/BOLD scaling was larger for the 1D as compared to 2D task in rostral regions (RLPFC, IFS, and Pre-PMd), but the reverse relationship held in the most caudal ROI, the PMd.

## Discussion

We show a paradoxical relationship between the BOLD response and RT during the performance of hierarchical tasks. While the IFS and PrePMd show an increased mean BOLD response to a more abstract "2D" task, consistent with prior work[3-5], they also show an increased scaling of BOLD with RT in a more concrete "1D" task. This difference in RT/BOLD scaling could not be explained as a result of regression dilution or differences in spectral frequency, and persisted even when 1D and 2D tasks were matched in terms of RT and all other experimental factors. The correlation with RT during the 1D task was not a global effect throughout the brain; the more caudal PMd, previously associated with more concrete tasks, instead showed greater RT/BOLD scaling during the more abstract 2D task in an FIR analysis.

This region- and task-specific double dissociation in the RT/BOLD scaling is incompatible with the view that both elevated RT and BOLD in hierarchical tasks arise solely from increased difficulty. If a single difficulty factor had caused both RT and BOLD change, then the RT/BOLD relationship should have been indistinguishable when these tasks were matched in terms of RT. We conclude that functional heterogeneity of these regions in lateral frontal cortex is unlikely to be wholly explained by a single "task difficulty" construct, at least as reflected in terms of time-on-task. These results specifically demonstrate this to be the case for rostrocaudal differences in prefrontal recruitment as a function of abstraction[2-5; c.f. 14].

If unitary task difficulty is insufficient to explain the pattern of functional data in frontal cortex, what does explain the differential scaling of the IFS and PrePMd with RT on the 1D task, even when RT was matched with the 2D task? We consider two possible accounts, both of which necessarily rely on the existence of functional specialization of lateral frontal cortex. One possibility is that subjects are actually engaging in higher-order control, of the type supported by these rostral regions, on these particularly slow trials of the 1D task. For example, though unnecessary, participants may occasionally treat a 1D task trial as though it were a 2D task trial. This would explain the functional specificity of this effect among both rostral and caudal frontal regions. In essence, this effect arises under conditions that do not require the function of the region in question. However, though plausible, we found no clear evidence to support any specific strategies being consistently used on slow trials in secondary analyses. For example, we did not find evidence that participants were unnecessarily encoding color as a higher order dimension on slow 1D trials (see Supplementary Text S1). Nevertheless, it is conceivable that these slow trials reflect a variable range of inefficient control strategies or other demands that recruit available capacity, supported by more rostral lateral frontal regions. Our approach would be insensitive to such a possibility.

A related alternative is that BOLD variance in areas supporting hierarchical control may be "quenched" by demands on hierarchical control. In that case, RT/BOLD scaling would necessarily be reduced in these areas, even if RTs were matched between tasks differing in hierarchical control demands. This follows observations in posterior cortex that stimuli matching regional tuning properties elicit a reduction in spiking variance, even as mean spike rates are increased[31-32]. Our results may reflect a fundamentally similar mechanism operating within frontal cortex. By this account, the hierarchical control demands of the 2D task differentially match the tuning properties of rostral regions of frontal cortex, and those of the more concrete 1D task differentially match the tuning properties of caudal areas. This matching quenches variance in these regions in a task-selective way. For illustration, consider the PrePMd, which appears responsive to demands on second-order hierarchical control. When those demands are encountered, a large set of neurons within the PrePMd would be expected to increase their activity, and hence BOLD. Any neurons in the prePMd with dissimilar tuning properties would, by contrast, be suppressed. Both effects should be reliable across time, and hence drive reduced variance in BOLD when second-order control demands are encountered. As a result, RT/BOLD scaling would be decreased in any region exhibiting both time-on-task effects *and* tuning properties that match the demands of the task itself, a possibility for which we also found some support (Supplementary Text S2 and Figure S1).

Our work significantly advances prior investigations of time-on-task effects in several ways[24-29; 33-37]. First, we identify methodological issues confronting the investigation of time-on-task effects[22-23], and adopt novel methods to resolve these issues. Second, we apply these methods in the domain of cognitive control, where recent theoretical and empirical work has suggested a potentially prominent role for a unitary difficulty construct[14-17; 24-29]. And, finally, we extend classical dissociation logic (and subsequent work on state-trace methodology) by allowing for non-monotonicity in how a single latent process maps to either of two dependent measures (Fig 1B-C).

However, one important caveat to our work comes from a related limitation inherent to classical dissociation logic. In prior applications of dissociation logic and state trace methods, latent cognitive processes must map monotonically all to dependent variables to draw valid inferences regarding the number of underlying processes[18-20]. Here, by contrast, we require only a monotonic mapping between the latent process and one dependent measure. This requirement is illustrated in Figure 8 (see also Figure 1). Even if the mapping from difficulty to RT is nonlinear, and the mapping from difficulty to BOLD is nonmonotonic (Figure 8A), matching two tasks on RT leads to the unavoidable prediction of an identical BOLD/RT relationship if a single underlying construct (e.g., "difficulty") is taken to drive these patterns (Figure 8B). However, if the mapping from difficulty to RT becomes nonmonotonic (Figure 8C), then RTs can appear to be matched while actually having distinct values of difficulty (Figure 8D).

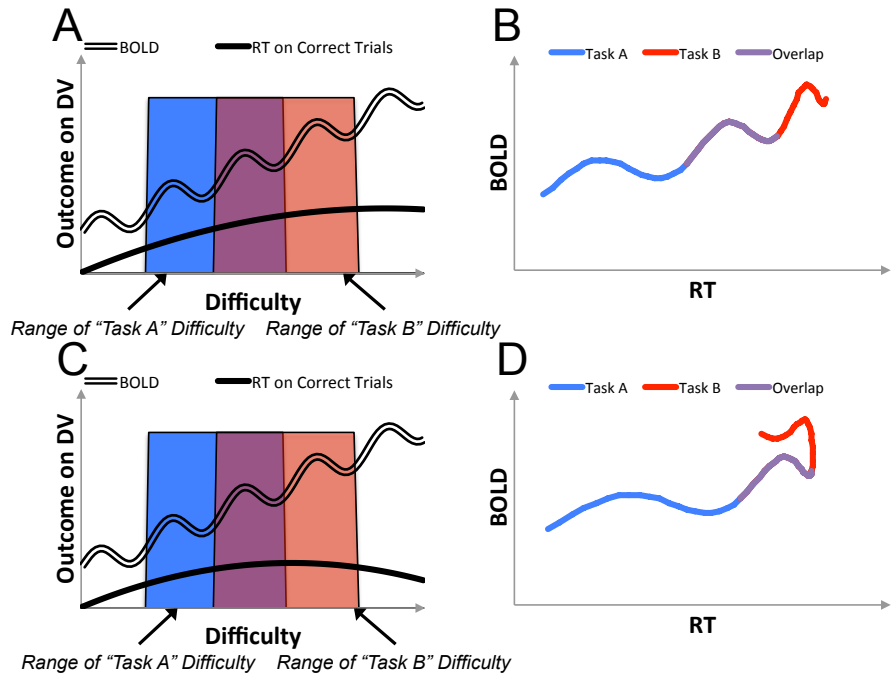

**Figure 8. Nonmonotonicity in the mapping of a single process to the matched dependent measure subverts the analysis of interdependent variables. A.** Here, both BOLD and RT are non-linear functions of difficulty. BOLD is also a nonmonotonic function of difficulty. **B.** Given the differences in difficulty between the two tasks illustrated in **A**, there are clear differences in their mean BOLD and RT. Nonetheless, when the tasks are matched on RT (purple region), the RT/BOLD relationship is necessarily the same across them. **C**. However, if both dependent measures are nonmonotonically mapped to difficulty, then it is possible to match the tasks on any one measure and still have mismatched difficulty. **D**. As a result, the RT/BOLD relationship may differ between tasks, even when RTs are matched between them.

In the current study, this requirement amounts to the assumption that difficulty is a monotonic function of RTs within the range of our matched distribution (600-1300ms). Two kinds of potential departures from non-monotonicity are worth considering. First, very long RTs might sometimes occur on trials where experienced difficulty was minimal, but performance was nevertheless delayed (e.g., due to inattention). Second, very short RTs might sometimes occur on trials where experienced difficulty was maximal, and so participants issue a quick guess rather than expending the mental effort to generate a response. Findings from the current study speak against both these possibilities. First, 2D trials falling in the lower half of the matched RT distribution (600-950ms) were in fact less error-prone than those in the upper half (950-1300ms; 2% vs. 8% errors, respectively; t(21)=2.56, p<.02), contrary to what would be expected if guessing were disproportionately common on trials with short RTs. Second, disproportionate mind-wandering on slow trials would predict greater BOLD in default mode, not frontoparietal regions in the slow 1D task trials, as compared to the fast 2D trials. Again, if anything, we observe the opposite pattern, with greater activation in default regions during the 1D relative to 2D task (Fig 4A). Thus, there is no evidence that non-monotonic difficulty effects could explain the patterns we report.

While our results therefore suggest a unitary construct of task difficulty cannot fully explain both RT and BOLD phenomena from hierarchical control tasks, these results should also not be taken to indicate that general difficulty makes no contribution in cognition or is not represented by the brain in any context. To the contrary, theoretical arguments for the importance of difficulty representations in domains such as mental effort and decision making are compelling[15-17]. Instead, we suggest that investigations of task-induced changes in RT/BOLD scaling will be important for validating a domain-general difficulty construct. For example, future work investigating the behavioral and neural correlates of a unified difficulty construct may explicitly posit mappings between this construct and both RT and BOLD, and search for regions where the implicitly predicted RT/BOLD relationship can be shown to be task-general, matching for any differences in the reliability, range, and spectral frequency of RT that escape experimental control.

A further methodological consideration raised by our results is that if RT/BOLD scaling is generally reduced in regions whose tuning properties match the demands of a given task[e.g. 31-32], then functional localization via trial-by-trial RT regressors could yield potentially counterintuitive results. Specifically, parameter estimates for trial-by-trial RT regressors might actually be higher in regions that are less associated with the task itself (the same would hold for estimates of percent signal change, even if variance-normalized statistics, like *z* statistics, are lower). This in turn motivates additional caution in comparing the neural correlates of individual differences in task performance with the neural correlates of trial-by-trial fluctuations in task performance.

Finally, our work highlights the potential importance of exploring how variance in BOLD signal intensity, rather than just its mean, may be influenced by experimental and behavioral factors[31-32; 38-39], beyond merely reflecting nuisance variance[40]. For example, a task may reduce BOLD variance because the task drives greater homogeneity in the response across multiple instances of the same event. Supplementary analysis provided preliminary evidence of such increased variance in prePMd, relative to PMd in, the 1D task relative to the 2D task (see Supplementary Text S2). Typical univariate analyses and also standard MVPA methods may be relatively insensitive to the presence of such higher-order moments in the BOLD timeseries[41]. Future methodological work may be needed to consider ways to better address these higher order effects[e.g. 42].

To conclude, by demonstrating this paradoxical relationship between BOLD and RT, we provide evidence that task difficulty does not yield a pattern in RT and BOLD across regions of lateral frontal cortex consistent with a unitary difficulty factor. Though task difficulty can be a useful heuristic for predicting when cognitive control or lateral frontal cortex is needed for task performance, regional involvement in these tasks appears driven by specific task demands, like the degree of abstraction, that have a more complex relationship with experienced difficulty.

**Online Methods**

*Participants*
Twenty-two right-handed adults (aged 18-35; 14 female) with normal or corrected-to-normal vision completed the experiment. All spoke English natively, were screened for contraindications for MRI as well as for the use of neurological medications and conditions, and provided informed consent in accordance with the Research Protections Office at Brown University.

*Task Design*

Training Phase. Subjects were trained in the scanner to press one of four buttons to each of four abstract shapes, and to each of four abstract textures. This training was administered in blocks of 35 trials for each dimension (shape vs. texture). Two blocks appeared per run, with a total of 8 runs of training. The correct feature-to-response mappings remained consistent throughout the entirety of the experiment.

Practice Phase. Following this training, subjects were given two (unscanned) blocks of 12 trials each in which to practice the experimental version of the tasks. Each block began with the presentation of the color-to-dimension mappings relevant for that block; only two colors were ever presented in a block. In "1D" blocks, both presented colors were associated with a single dimension (shape or texture). In "2D" blocks, one color was associated with shape, and the other color with texture. Like the feature-to-response mappings, these color-to-dimension mappings also remained consistent throughout the task. Thus, subjects were required to attend to the color surrounding each stimulus only in the 2D blocks.

Testing Phase. Following these 24 practice trials, subjects completed six scanned runs consisting of 48 trials each. Each run consisted of two blocks of each task, occurring in counterbalanced order. As in the practice, each task occurred in a block of 12 consecutive trials, and began with the presentation of an instruction screen for 16 seconds showing the color-to-dimension mappings relevant for that block. Four of the 16 possible shape x texture combinations were presented with 70% frequency; the remaining 12 stimuli, corresponding to the other possible feature combinations, were presented with only 30% frequency. This frequency manipulation, while controlled for in all analyses presented here, is not of primary interest in the present report.

*MRI procedure*

During training and experimental phases of the task, whole-brain imaging was performed with a Siemens 3T TIM Trio MRI system using a 32-channel head coil. A high resolution T1 MPRAGE was collected from each participant at the beginning of each session. Each of the six runs of the experimental task involved 128 functional volumes, with a fat-saturated gradient-echo echo-planar sequence (TR = 2s, TE=28ms, flip angle = 90°, 33 interleaved axial slices, 192mm FOV with voxel size of 3x3x3.5mm). Head motion was restricted with padding, visual stimuli were rear-projected and viewed with a mirror attached to the head coil.

*Data preprocessing*
*Behavioral data.* Reaction times less than 250ms or more than 2s were excluded from analysis, resulting in an exclusion of less than 3.5% of all trials; all reaction time effects are reported for correct trials, except where noted. Statistical analysis of error rates was conducted after variance-stabilizing arcsin-square root transform.

*Imaging data.* Data were processed using a combination of SPM8 and FSL5.1.0. First, SPM8 tools (artglobal and tsdiffana) were used for artifact detection, and slice timing correction was then performed. Data were motion-corrected using rigid transformations in MCFLIRT to the middle acquisition of each run. Runs with movement of more than 2mm were excluded from analysis, resulting in the exclusion of 3 runs total (one from each of 3 subjects). Four additional subjects were excluded from analysis due to >2mm translation in more than one run. Grand-mean intensity normalization of the entire 4D dataset was performed with a single multiplicative factor, and the data were subjected to a temporal highpass filter (Gaussian-weighted least-squares straight line fitting, with sigma=100s); the data were then smoothed at 8mm FWHM. The middle acquisition of each run was then registered to each participant's brain-extracted MPRAGE, via boundary-based registration, with a linear 7DOF transform, and the MPRAGE was registered to the MNI standard brain using a linear 12DOF transform.

*Statistical Analysis*
   GLMs were estimated using FEAT (FMRI Expert Analysis Tool) version 5.98 (FMRIB's Software Library, www.fmrib.ox.ac.uk/fsl), on the basis of explanatory variables (EVs) coding for all possible conjunctions of the following event types: task [D1 vs. D2], accuracy [correct vs. incorrect], and frequency [high vs. low]. In addition, contrast-coded regressors were estimated for task-specific effects of stimulus congruency [congruent vs. incongruent], dimension switches [switch vs. repeat of color], and response switches [switch vs. repeat of the specific button press]. These additional regressors were included in a "maximal model," as noted in the text, to help ensure the effects of interest could not be attributed to any of a variety of other experimental factors. The duration of all aforementioned events was uniformly set to .5s. Task-specific sustained regressors were also used in the maximal model; the duration of these sustained regressors was made equal to the duration of each task block.

   For models investigating time-on-task, additional parametric EVs were added to these models. These EVs were duration-modulated (aka the "variable-epoch model"), given the increased power of such regressors for detecting RT-related BOLD responses, and the increased specificity to time-on-task effects conferred by including them in models already containing constant epoch regressors[37]. The duration of each such event was set to the actual duration of each trial's RT. As noted in the text, duration-modulated regressors were constructed with separate EVs for 1D and 2D trials independently.

In all models, EVs corresponding to the 6 degrees of motion estimated by MCFLIRT, and transient responses to the instruction events, were entered as EVs of no interest. Likewise, in all models, all EVs except those corresponding to motion were convolved with a standard double-gamma hemodynamic response function (HRF), high-pass filtered in the same way as the functional data, and then used as a regressor in the GLM and FSL's prewhitening tool, FILM. Results presented in the main text are from models that include temporal derivatives, however all Results were robust when temporal derivatives were omitted, suggesting that any variance attributed to these derivatives does not alter the conclusions presented here.

Second-level analyses were constructed for each subject by considering runs as a fixed effect. Group-level whole-brain analyses were conducted with FLAME by considering subjects as a random effect. We used a voxel-level threshold of z>2.58 and a further cluster-based family-wise error rate correction to p<.05 using FSL's implementation of Gaussian Random Field theory, which adopts contrast-specific extent thresholds given each contrast's effective resolution (resels).

For FIR models, we estimated 10 beta-coefficients corresponding to the effect of z-transformed RT across each of the 10 volumes following trial onset[36]. These analyses represent a model-free method of interrogating any changes in the amplitude and shape of the hemodynamic response that may scale with RT. The FIR parameter estimates were averaged across hemispheres to assess BOLD percent signal change (PSC) as a function of RT in each bilateral ROI. Owing to the larger number of parameters that must be estimated for a FIR model, it was not possible to fit them for the larger "maximal model."

Matching analyses were conducted using the genmatch procedure of the Matching package version 4.8-3.4[43]. This procedure uses a genetic algorithm to find one-to-one matches between trials from each tasks, so as to maximize the multivariate balance between tasks in terms of RT, run, frequency, congruency, switching, and their interactions on correct trials. Matching was conducted without replacement and a caliper of .15. The genetic algorithm employed a population of 2500 individuals and 1000 generations, and stopped only when 5 consecutive generations yielded no change in the identity of included trials. This procedure was carried out for each subject independently, ensuring both within- and between-subject multivariate matching of trials between the 1D and 2D tasks. It yielded an average of 22 matched trials for each task per subject (range: 15-28). Runs in which no trials could be matched across tasks were dropped from analyses involving matched trials (n=4).

All statistical tests are two-tailed, with corrected alphas of .05.

**Acknowledgements**

We thank our reviewers and funding agencies. This study was supported by awards from the National Institute of Neurological Disease and Stroke (R01 NS065046), the Alfred P. Sloan Foundation, and the James S. McDonnell Foundation.

**Supplementary Information**

Supplementary Table 1: Peak Coordinates from Maximal Model Contrasts of Interest

| Contrast | Label | Peak (MNI) | | | Peak Z | Extent (voxels) |
|---|---|---|---|---|---|---|
| | | X | Y | Z | | |
| D2>D1 Task Events | LOC | -24 | -64 | 44 | 4.72 | 12981 |
| | SFG | -22 | 8 | 56 | 4.64 | (local max) |
| | MFG | -46 | 10 | 32 | 4.61 | (local max) |
| | SPL | -30 | -54 | 60 | 4.59 | (local max) |
| | Insula | 32 | 24 | 0 | 4.42 | (local max) |
| | PCG | -4 | 16 | 50 | 4.41 | (local max) |
| | O Fusiform | 16 | -72 | -18 | 3.96 | 621 |
| | FPC | -32 | 56 | 6 | 4.02 | 561 |
| D1 Task Events | Occipital Pole | 32 | -90 | 4 | 6.11 | 16760 |
| | LOC | 42 | -86 | 10 | 5.86 | (local max) |
| | LOC | 38 | -90 | 0 | 5.73 | (local max) |
| | Occipital Pole | -12 | -102 | 12 | 5.71 | (local max) |
| | O Fusiform | 38 | -64 | -10 | 5.63 | (local max) |
| | TPJ | 36 | -56 | -14 | 5.59 | (local max) |
| | Post-C Gyrus | -52 | -14 | 44 | 4.08 | 1792 |
| | Post-C Gyrus | -48 | -16 | 44 | 4.05 | (local max) |
| | Post-C Gyrus | -56 | -14 | 52 | 3.98 | (local max) |
| | Pre-C Gyrus | -62 | 2 | 30 | 3.78 | (local max) |
| D2 Task Events | TO Fusiform | 30 | -52 | -14 | 7.57 | 64727 |
| | SPL | -28 | -56 | 58 | 7.1 | (local max) |
| | Post-C Gyrus | -42 | -28 | 54 | 7.06 | (local max) |
| | TO Fusiform | 25 | -58 | -18 | 6.95 | (local max) |
| | O Fusiform | 32 | -62 | -14 | 6.9 | (local max) |
| | O Fusiform | 34 | -80 | -10 | 6.89 | (local max) |
| Effect of D1>D2TOT | SMG | -48 | -40 | 52 | 4.12 | 3466 |
| | SMG | -48 | -34 | 42 | 3.77 | (local max) |
| | Precuneous | -10 | -66 | 42 | 3.75 | (local max) |
| | Post-C Gyrus | -50 | -34 | 60 | 3.74 | (local max) |
| | LOC | -26 | -60 | 48 | 3.68 | (local max) |
| | SPL | -28 | -48 | 54 | 3.61 | (local max) |
| | Pre-C Gyrus | -40 | 4 | 40 | 3.96 | 1582 |
| | MFG | -40 | 32 | 28 | 3.73 | (local max) |
| | MFG | -54 | 16 | 32 | 3.57 | (local max) |
| | IFG | -34 | 18 | 26 | 3.49 | (local max) |
| | MFG | -28 | -4 | 52 | 3.26 | (local max) |
| | Pre-C Gyrus | -48 | 6 | 24 | 3.22 | (local max) |
| | MFG | 50 | 12 | 36 | 3.79 | 1388 |
| | MFG | 40 | 10 | 32 | 3.54 | (local max) |
| | MFG | 40 | 30 | 26 | 3.54 | (local max) |
| | Pre-C Gyrus | 40 | -2 | 38 | 3.51 | (local max) |
| | MFG | 44 | 26 | 34 | 3.48 | (local max) |
| | SFG | 22 | -6 | 54 | 3.42 | (local max) |
| | FPC | -30 | 52 | 14 | 3.79 | 599 |
| D1 TOT Effect | LOC | -8 | -72 | 62 | 4.31 | 6406 |
| | LOC | -28 | -68 | 50 | 4.08 | (local max) |
| | LOC | -10 | -68 | 64 | 4.06 | (local max) |
| | LOC | -32 | -60 | 48 | 3.94 | (local max) |
| | SPL | -24 | -58 | 50 | 3.92 | (local max) |

RUNNING HEAD: Rostrocaudal reaction time effects22|  | Region | x | y | z | Z | Cluster size |
|---|---|---|---|---|---|---|
|  | Precuneous | -4 | -70 | 50 | 3.9 | (local max) |
|  | SFG | -44 | 34 | 46 | 4.33 | 3699 |
|  | MFG | -34 | 16 | 28 | 3.84 | (local max) |
|  | MFG | -48 | 16 | 32 | 3.83 | (local max) |
|  | MFG | -56 | 16 | 32 | 3.8 | (local max) |
|  | MFG | -40 | 28 | 28 | 3.78 | (local max) |
|  | MFG | -48 | 8 | 54 | 3.75 | (local max) |
|  | MFG | 50 | 24 | 34 | 4.19 | 2674 |
|  | MFG | 34 | 0 | 44 | 3.86 | (local max) |
|  | MFG | 32 | 2 | 50 | 3.83 | (local max) |
|  | Caudate | 20 | 28 | -6 | 3.64 | (local max) |
|  | MFG | 42 | 32 | 30 | 3.5 | (local max) |
|  | SFG | 32 | 8 | 70 | 3.46 | (local max) |
|  | ITG | -48 | -58 | -8 | 3.85 | 1173 |
|  | O Fusiform | -42 | -68 | -18 | 3.71 | (local max) |
|  | LOC | -48 | -70 | -8 | 3.66 | (local max) |
|  | ITG | -50 | -56 | -14 | 3.59 | (local max) |
|  | MTG | -56 | -60 | 2 | 3.46 | (local max) |
|  | TO Fusiform | -34 | -48 | -24 | 2.96 | (local max) |
|  | PCG | -6 | 14 | 52 | 3.62 | 953 |
|  | FPC | -40 | 64 | 10 | 3.7 | 801 |
| D2 TOT Effect | Occipital Pole | -30 | -96 | -10 | 4.49 | 5399 |
|  | LOC | 32 | -78 | 0 | 4.25 | (local max) |
|  | O Fusiform | 18 | -86 | -8 | 4.1 | (local max) |
|  | Occipital Pole | 22 | -94 | 22 | 4.1 | (local max) |
|  | Occipital Pole | 10 | -100 | 16 | 4.02 | (local max) |
|  | LOC | 38 | -84 | 2 | 4 | (local max) |
|  | Putamen | 26 | -30 | 8 | 4.41 | 723 |

Supplementary Text S1: *No Evidence for Hierarchical Control on Slow 1D Task Trials*

To assess whether subjects may have been engaging in hierarchical control on the slow trials of 1D task – for example, by erroneously attending to color – we used multivariate pattern analysis (PyMVPA; Hanke, et al. 2009).

We reasoned that if subjects were erroneously attending to color on slow 1D trials, then color should be more accurately classified on slow, as compared to fast 1D trials. To test this hypothesis we first extracted the motion-corrected timeseries from each voxel within each of our a priori ROIs, and divided them into "chunks" defined by run for cross-validation. Each timeseries was linearly detrended, and z-scored with respect to rest. The detrended and z-scored BOLD intensities occurring 4-6 seconds following each 1D stimulus were then classified according to the color used on that trial, using a linear support vector machine (LinearCSVMC in PyMVPA). Separate classifiers were trained and tested on 1D trials falling above and below each subject's mean RT. In neither "fast" nor "slow" 1D trials did we observe classification accuracy exceeding that observed across 1000 permuted training sets; if anything, color-decoding accuracy relative to the permuted null was somewhat higher on short as compared to long trials (ranking in the top 36.5% vs. 39.6% of the permuted distribution). Similar results were achieved with many variants on this procedure, including the provision of BOLD in each of the first 3 post-stimulus TRs to the classifier, feature selection using one-way ANOVA, subsampling of the data to ensure balanced classes, and z-scoring voxels within ROIs for each time-point separately.

As a second attempt, we trained classifiers on 2D task trials, and assessed their generalization to 1D task trials as a function of RT. In none of our a priori ROIs did we observe a significant correlation between classifier accuracy and RT (mean R=.04, thus accounting for less than 1% of the variance in RT). As above, many variants on this scheme were assessed with no qualitative change to the results.

Supplementary Text S2: *Variance Analyses As a Function of Task and Area*

If the demands of the 2D task better match the tuning properties of the PrePMd than the 1D task, intrinsic variability in the PrePMd may be quenched by the 2D task, and not the 1D task. This would translate to greater heterogeneity in the response of the PrePMd across trials of the 1D task, as compared to the 2D task, and as compared to the PMd (see Discussion). To test this hypothesis, we used a robust measure of variability (the median absolute deviation; MAD[44]) in the residuals of the models assessing trials matched for RT across the 1D and 2D task. The ToT regressor was removed from these models for this analysis, to reveal variability which the ToT regressor could explain in principle. We then averaged the MAD across the voxels of each volume acquired 2-8s following the onset of each matched trial.

This analysis revealed that the PrePMd shows larger residual variability from 2-8s post-stimulus on matched trials of the 1D task, relative to the matched trials of the 2D task ($t(21)=3.26$, $p=.004$). No such pattern was observed within PMd ($t(21)=1.26$, $p>.22$). The interaction with ROI is significant ($F(1,21)=6.96$, $p=.015$; See Figure S1A). Even when analyzing the residuals of these same trials in the "maximal" model – that is, the model controlling for the effects of congruency, dimension switches, and response switches – the same interaction holds (Figure 1B; $F(1,21)=5.24$, $p=.03$; note however that neither simple effect remains significant in this analysis, p's>.24). These effects were observed only around the anticipated peak of the HRF (e.g., p's >.15 for the period 0-4s prior to matched trial onset, or from 10-20s after matched trial onset).

These results are consistent with the notion that the 1D task does not match the tuning properties of the PrePMd as well as the 2D task. As a result, the 1D task may less effectively "quench" the intrinsic variability of the PrePMd response (see Discussion). This greater residual variability in the BOLD response in the PrePMd could therefore enable a larger scaling of BOLD with factors that vary from trial-to-trial.

Supplementary Figure S1

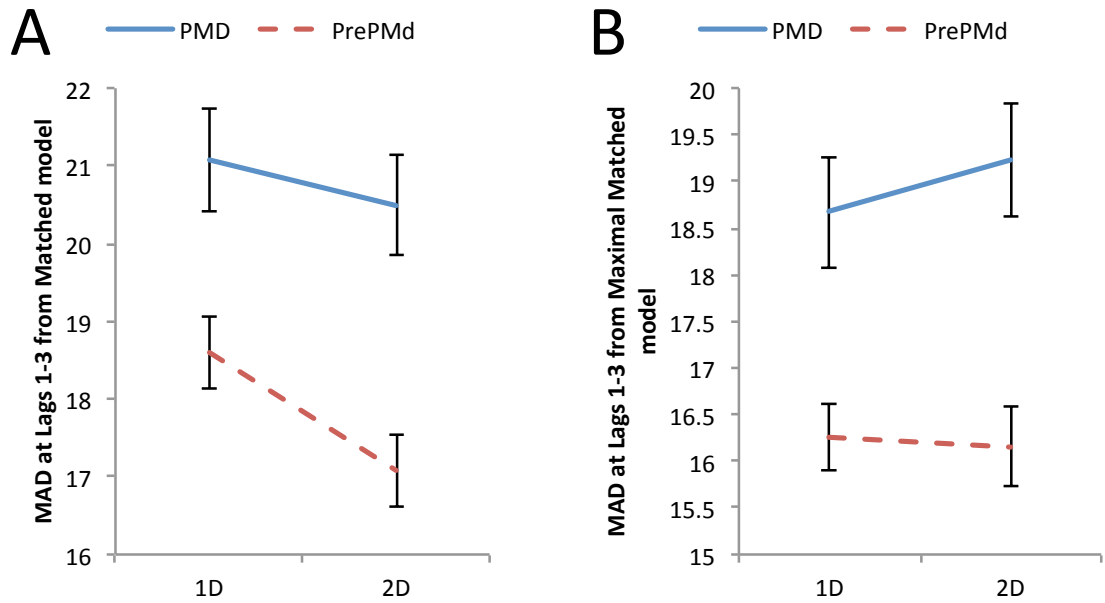

**Figure S1 – The median absolute deviation (MAD) of the residuals on RT-matched trials in the 1D and 2D tasks, as a function of area [PrePMd vs. PMd], in the standard (A) and maximal models (B) when the time on task regressor was omitted. A**. Relative to the PMd, the PrePMd shows disproportionately greater residual variability in the 2-8s following matched trials of the 1D, as compared to 2D task. **B.** Even when additionally controlling for a number of other experimental factors, the PMd and PrePMd show discrepant patterns in residual variability as a function of task.